\documentclass[12pt]{article}
\usepackage{amssymb,latexsym}
\usepackage[mathscr]{eucal}

\newcommand{\mmm}{m_c}
\newcommand{\cM}{{\cal M}}
\newcommand{\cN}{{\cal N}}
\newcommand{\vphi}{\varphi}

\newcommand{\g}{{\mathfrak g}}

\newcommand{\F}{{\mathcal F}}

\newcommand{\Z}{{\mathbb Z}}
\newcommand{\R}{{\mathbb R}}
\newcommand{\C}{{\mathbb C}}

\newcommand{\beq}{\begin{equation}}
\newcommand{\eeq}{\end{equation}}
\newcommand{\beqa}{\begin{eqnarray}}
\newcommand{\eeqa}{\end{eqnarray}}

\newcommand{\tr}{\mbox{\rm tr}\,}
\newcommand{\uno}{\mbox{1\kern-.59em {\rm l}}}

\newcommand{\nn}{\nonumber}
\newcommand{\be}{\begin{equation}}
\newcommand{\ee}{\end{equation}}
\newcommand{\bea}{\begin{eqnarray}}
\newcommand{\eea}{\end{eqnarray}}

\newcommand{\ft}[2]{{\textstyle\frac{#1}{#2}}}

\newcommand{\tinyyoung}[1]{\mbox{\tiny\young(#1)}}
\usepackage[vcentermath,enableskew]{youngtab}
\begin{document}
\begin{titlepage}
\begin{flushright}
{ROM2F/2005/21}\\
{SISSA 73/2005/FM}\\
{CERN-TH/2005-198}\\
\end{flushright}
\begin{center}

{\large \sc ${\cal N}=1$
 Superpotentials from
Multi-Instanton Calculus}\\

\vspace{0.2cm}
{\sc Francesco Fucito}\\
{\sl Dipartimento di Fisica, Universit\'a di Roma ``Tor Vergata''\\
I.N.F.N. Sezione di Roma II,\\
Via della Ricerca Scientifica, 00133 Roma, Italy}\\
\vskip 0.2cm
{\sc Jose F. Morales}\\
{\sl CERN, TH Division\\
1211 Geneva 23, Switzerland}\\
\vskip 0.2cm
{\sc Rubik Poghossian}\\
{\sl Yerevan Physics Institute\\
Alikhanian Br. st. 2, 375036 Yerevan, Armenia}\\
and\\
{\sc Alessandro Tanzini}\\
{\sl Scuola Internazionale Superiore di Studi Avanzati\\ I.N.F.N. Sez. di Trieste\\
Via Beirut 4, 34014 Trieste, Italy} \\
\end{center}
\vskip 0.5cm
\begin{center}
{\large \bf Abstract}
\end{center}
{ In this paper we compute gaugino and scalar condensates in
${\cal N}=1$ supersymmetric gauge theories
with and without massive adjoint  matter,
using localization formulae over the multi--instanton moduli space.
Furthermore we compute the chiral ring relations among the correlators
of the ${\cal N}=1^*$ theory and check this result against
the multi-instanton computation finding agreement.}

\par    \vfill
\end{titlepage}
\addtolength{\baselineskip}{0.3\baselineskip}
\setcounter{section}{0}
\section{Introduction}
\setcounter{equation}{0}
The vacua of a non-abelian supersymmetric gauge theory (SYM) look very different according to
the number of supersymmetries (SUSY)\cite{Witten:1982df}. While in the ${\cal N}=1$ case a theory with gauge
group $SU(N)$ has $N$ vacuum states, for extended supersymmetry
it happens that the theory has flat directions.
If SUSY is unbroken, the classical degeneracy is not lifted by quantum corrections. The vacua of the
quantum theory then form a moduli space. Computations of non-perturbative effects must
cope with these facts.

The strategy developed in \cite{Shifman} for the computation of correlators
in ${\cal N}=1$ SYM with massive matter, $m$, in the fundamental representation is simple:
first perform the computation
in the presence of a generic vacuum expectation value (v.e.v.), $a$, for the
scalar fields, then determine the v.e.v. by minimizing the quantum potential.
In the lectures in Ref.\cite{Shifman:1997nw} a nice description of this procedure
is given. Let us focus on the computation of the quantum potential.
In this example since all the gauge degrees of
freedom are massive due to the Higgs mechanism, the potential can only be
a function of the matter fields. By using $R$--symmetry arguments, in the $SU(2)$ case one finds
a quantum $F$-term $F=ma-2\frac{\Lambda^5}{a^3}$.
Minimizing this potential one finds $a^2=\pm\sqrt{2}\Lambda^{5/2}/\sqrt{m}$
in agreement with the previous statements concerning the number of vacua of a ${\cal N}=1$ SUSY theory.
To generalize this procedure to other types of massive matter is non-trivial: for matter in the adjoint
there are massless gauge degrees of freedom and $R$--symmetry arguments alone do not
determine the quantum potential. But the findings of Ref.\cite{sw} can help in this respect: the
structure of the quantum effective theory studied in this reference is coded in the geometry of the
Seiberg-Witten curve. Even more interestingly a picture of confinement is obtained when the curve
degenerates at some special points where the monopoles are massless. These degenerate points lie in
the moduli space of SYM and a connection with the v.e.v.'s of the previous discussion is not straightforward.

In this paper we will show how the dramatic progress in the understanding of
multi-instanton corrections to SYM with extended SUSY, recently
achieved \cite{Nekrasov:2002qd,Flume:2002az,Bruzzo:2002xf,Losev:2003py,Nekrasov:2003rj,
Flume:2004rp,Marino:2004cn,Nekrasov:2004vw,Fucito:2004gi}, can be used to provide the missing link.
In turn, this will lead to the application of these techniques to ${\cal N}=1$ SYM. In particular we
will focus on ${\cal N}=1$ SYM  obtained
by mass deformations of ${\cal N}=2$ and ${\cal N}=2^*$ SYM theories.  The deformations lift
the moduli space degeneracy and the ${\cal N}=1$ correlators can be read from the ${\cal N}=2$
multi-instanton results once they are evaluated at the minima of the quantum potential.

This is the plan of the paper: in the first part of section 2 we discuss localization formulae in
presence of the chiral observables associated to the superpotential.
We show how to express the chiral condensates of the ${\cal N}=1$ theory in terms of
correlators of the parent ${\cal N}=2$ theory. The next task is to find the value of the v.e.v. of the scalar
fields in the ${\cal N}=1$ vacuum.

In section 3 we address this problem for ${\cal N}=1$ SYM with an adjoint chiral multiplet and a non-trivial
superpotential or mass. We complement the discussion
with the results of \cite{sw}, where a vacuum showing the
signatures of confinement is exhibited. To do so one must first break ${\cal N}=2\to {\cal N}=1$ SYM
by adding a mass.
In this vacuum the monopoles are massless
and this condition takes a nice geometrical form: it corresponds to the
degeneration of a Riemann surface in which the B-cycles are zero.
The variable $u(a)=\langle\tr\phi^2\rangle$ takes a definite value at
this point. Inverting the series $u(a)$ gives the value of the v.e.v.'s
 in the minimum of the quantum
potential. The corresponding condensates that we find
via multi--instanton calculus
are then easily checked against chiral ring relations along the same lines of
\cite{Flume:2004rp}.

In section 4 we discuss the ${\cal N}=1^*$ case.
The situation here is more involved. The results
of \cite{Seiberg:1994aj,Donagi:1995cf} are not enough to carry
our program out. So we supplement them
by computing chiral ring type equations and using the connection between
the ${\cal N}=1^*$ SYM and
integrable models \cite{Gorsky:1995zq,Martinec:1995by}.
Also in this case we check the results of multi--instanton calculus
against chiral ring type relations.

Finally in section 5 we summarize our results.

\section{Localization formulae for ${\cal N}=1$ supersymmetric theories}
\setcounter{equation}{0}
\subsection{Equivariant forms}
Let $\g=U(1)^r$ be a group action on a smooth manifold $M$ of complex dimension $\ell$
specified by the vector field
$$
\xi=\xi^i(x){\partial\over \partial x^i}
$$
Let $\Omega(M)$ be the space of differential forms on $M$.
The equivariant extension is given by the polynomial map
$\alpha:\g\rightarrow \Omega(M)$ from $\g$ to the algebraof differential forms on $M$. $\alpha$ may be regarded as an element of
$\C[\g]\otimes\Omega(M)$ with $\C[\g]$  the algebra
of complex-valued polynomials on $\g$.
We define a grading in $\C[\g]\otimes\Omega(M)$ by letting, for homogeneous
$P\in\C[\g]$ and $\beta\in\Omega(M)$,
\beq
\deg(P\otimes\beta)=2 \deg(P)+\deg(\beta).
\label{deg}
\eeq
Let us then introduce the equivariant derivative, $d_\xi$, and the Lie derivative, ${\cal L}_\xi$,
$$
d_\xi \equiv d-i_\xi ,  \quad\quad
d_\xi^2= - d i_\xi - i_\xi d= - {\cal L}_\xi
$$
where $d$ is the exterior derivative and $i_\xi dx^i\equiv \delta_\xi x^i$ the contraction with $\xi$.
If $\alpha$ satisfies
 \be
 d_\xi \, \alpha=0
 \ee
it is said to be equivariantly closed.
Let $x^s_0$ be the isolated critical points of the group action $\xi$
i.e. points where $\xi^i(x^s_0)=0\, ,\forall i$. The integral of an
equivariantly closed form is given by the localization
formula
\be
\int_M\, \alpha=(-2\pi)^\ell\,\sum_{s}\,
{\alpha_0(x_0^s)\over {\rm det}^{1\over 2} {\cal L}_\xi(x^s_0)}
\label{loc}
\ee
with ${\cal L}_{\xi}{}_i^j=\partial_i \xi^j: T_{x_0} M\to T_{x_0} M$
the tangent space map induced by the vector field $\xi$.
$\alpha_0$ is the zero form part with respect to the grading in $\Omega(M)$.

The localization formula (\ref{loc}) is an extremely powerful tool
to compute integrals as for example volumes and Chern classes
of complicate Riemannian manifolds
(in our case the ADHM moduli space).
As an example let us compute the integral of the form $e^{-a(x^2+y^2)}\, dx\wedge dy$
via $U(1)$ localization on $\R^2$. We choose $\xi$ to be the generator of $SO(2)$ rotations
in the plane: $\xi =\hbar \, (y\,\partial/\partial x - x \,\partial/\partial y)$
with critical point $x_0=y_0=0$.
The equivariant extension of the form given above is
\be
\alpha=e^{-a (x^2+ y^2)}\,dx\wedge dy-{\hbar\over 2 a}e^{-a(x^2+y^2)}=\alpha_2+\alpha_0
\ee
With respect to the grading (\ref{deg}), $\alpha$ is a two-form which is made of two
parts having respectively degree two and zero
in $\Omega(M)$.
Sending $\hbar\to 0$ we recover the starting form.
Using the localization formula we get
\be
\int_{\R^2}\, e^{-a(x^2+y^2)}\,dx\wedge dy=
\int_{\R^2}\, \alpha=2\pi\,{ \hbar\, e^{-a(x_0^2+y_0^2)}\over 2 a \hbar}
={\pi\over a}
\ee
which is the result we would have obtained by performing the integral on the l.h.s.
Notice that the right
hand side does not depend on $\hbar$.
After this quick exercise we are ready to write down the
equivariant forms in $\C^2=\R^4$ under a $U(1)$ subgroup of $SO(4)$
generated by
$\xi = i\, \hbar \, (z^1 dz^1-z^2 dz^2-{\rm h.c.})$
\bea
\alpha_0 &=& 1 \nn\\
\alpha_{(2,0)}&=& dz^1\wedge dz^2+i\, \hbar\, z^1 z^2\nn\\
\bar{\alpha}_{(0,2)}&=& d\bar{z}^1\wedge d\bar{z}^2
-i\, \hbar\, \bar{z}^1 \bar{z}^2\nn\\
{\alpha}_{(2,2)}&=& dz^1\wedge dz^2\wedge d\bar{z}^1\wedge d\bar{z}^2+
i\, \hbar\,( z^1 z^2d\bar{z}^1\wedge d\bar{z}^2-{\rm h.c.})
- \hbar^2\, z^1 z^2\bar{z}^1 \bar{z}^2\label{forms}
\eea
The subscripts are the gradings with respect to the complex structure.
Notice that all the forms in (\ref{forms}) are invariant under the $U(1)$
action
\be
z^1\to e^{i\hbar} \, z^1, \quad z^2\to e^{-i \hbar} \, z^2
\ee
generated by $\xi$.

\subsection{Equivariant forms for supersymmetric theories}
Here we consider ${\cal N}=1$ SYM with gauge group $U(N)$ coupled to an adjoint
chiral superfield $\Phi$ with superpotential $W(\Phi)$
\be
S_{{\cal N}=1}=S_{{\cal N}=2}+ \int d^4x d^2\theta\, W(\Phi)
\label{n2w}
\ee
In the multi--instanton calculus, we treat the chiral tree-level
superpotential $W(\Phi)$ as the insertion of a set of non--trivial observables. In this way,
the correlators of the chiral ring of  ${\cal N}=1$ will be related to the correlators
in the underlying ${\cal N}=2$ SYM.
As it is well known, in a saddle point evaluation
of the functional integral with action $S_{{\cal N}=1}$ the quantum fluctuation of the fields
cancel among each other and one is left with the zero modes only.
Localization will then be performed in the ${\cal N}=2$ universal moduli space ${\cal M}\times \C^2$
where ${\cal M}$ is the moduli space of gauge connections. Our first task is to introduce equivariantly closed forms
suitable for being integrated with the localization formula. These closed forms will
generate the set of non--trivial observables. The equivariantly closed forms are given by a
combination of the zero modes of the theory suitably deformed by the presence of the torus action generated by $\xi$.
Let us introduce the equivariant derivative
\be
d_\xi=d_{\C^2}+d_{\cal M}-i_\xi
\ee
$d_{\C^2}$ is the usual total differential in $\C^2$, $d_{\cal M}$ is the total differential
on the moduli space of gauge connections\footnote{It is well known
that this space is obtained by starting from a flat space of real dimension $4k^2+4kN$,
subtracting $3k^2$ ADHM constraints and finally modding by the symmetry $U(k)$. For our purposes
it is enough to take $d_{\cal M}$ as the total differential of the space of real dimension
$4k^2+4kN$.} with winding number $k$
and $\xi$ is the fundamental vector field generating the action of the group
$ U(1)_{\hbar}\times U(1)^{N}_{a_\alpha}\times U(1)^k_{\phi}$.
 This group parameterizes the Cartan of the group of symmetries of the ADHM manifold:
the $SO(4)$ Lorentz group, the $U(k)_{\phi}$ symmetry
of the ADHM constraints
and the $U(N)_{a_\alpha}$ gauge symmetry. In an ${\cal N}=2$ theory the $a_\alpha$'s are free
parameters describing the expectation values of the scalar fields while in the ${\cal N}=1$
gauge theory under study here they will describe the couplings in the superpotential and the
${\cal N}=1$ vacuum.

Following \cite{Baulieu:1988xs}, let us introduce the universal equivariant
connection ${\cal A}$ and the universal equivariant field strength
${\cal F}$ (see Sect.7 of \cite{Baulieu:2005bs} for a detailed description of the geometrical framework).
Given the kernel $U$ of the ADHM matrix,\footnote{The idea is that a non-trivial bundle
of a certain dimension can always be mapped in a
trivial one of greater dimension. $U$ is the map between the two bundles and $P=UU^\dagger$ the projection
operator. It is then easier to find the covariant derivative on the non-trivial bundle as the
projection of the flat derivative of the trivial bundle i.e. $\nabla=P\partial$.} they can be written as
\bea
{\cal A}&=& \bar{U} \, d_\xi U=A+C\nn\\
{\cal F}&=& d_\xi \left(\bar{U} \, d_\xi U\right)
=F+\Psi+\varphi
\eea
in terms of the zero modes
\bea
A &=& \bar{U}d_{\C^2} U  \quad\quad  C=\bar{U} d_{\cal M} U\nn\\
F &=& F_{\mu\nu} dx^\mu dx^\nu=d_{\C^2}
\left(\bar{U} \, d_{\C^2} U\right)\nn\\
\Psi&=&\lambda_m dz^m +\psi_{\bar m} d\bar{z}^{\bar m}=
d_{\C^2} \left(\bar{U} \, d_{\cal M} U\right)+d_{\cal M}(\bar{U} \,
d_{\C^2} U)\nn\\
\varphi &=& (d_{\cal M}\bar{U}) (d_{\cal M} U)
-\bar{U}\, {\cal L}_\xi \, U\label{fpsi}
\eea
With respect to the grading of forms in the space $\C^2 \times {\cal M}$,
$A$ is a $(1,0)$ form, $C$ a $(0,1)$ form, $F$ is a $(2,0)$ form and $\Psi$ a $(1,1)$ form.
$U$ is a $(0,0)$ form and therefore $i_\xi U=0$. Notice that only $\varphi$ gets deformed
by $\xi$ as expected \cite{Flume:2004rp}.
In the limit $\hbar\to 0$ we recover the gauge connection,
field strength, gauginos and scalar field solutions in the instanton background.

Notice that in (\ref{fpsi}) we decomposed the real one--form $\Psi$
with respect to the complex structure of $\C^2$ as a complex $(1,0)$--form
$\lambda_m dz^m$ plus a complex $(0,1)$--form $\psi_{\bar m} d\bar{z}^{\bar m}$.
These two components can be identified respectively with the gaugino
and the chiral matter field of the ${\cal N}=1$ theory in the topologically twisted
formulation \cite{Johansen:1994aw,Witten:1994ev}. This identification allows us to write
the basic operators of the ${\cal N}=1$ theory in an equivariant form
in terms of ${\cal F}$ and the
"volume" forms (\ref{forms}) described in the last section \footnote{The authors of \cite{Flume:2004rp} have given
arguments for the closure of the form $\varphi$. Their reasonings should be replaced by the
present analysis.}
\bea
&& \int_{\R^4} d^4x\, {\rm tr}\, \varphi^k =\int_{\C^2} \alpha_{(2,2)}\wedge {\rm tr}\, {\cal F}^k
\nn\\
&& \int_{\R^4} d^4x\, {\rm tr}\, \lambda^\alpha \lambda_\alpha
\varphi^{k-2}  = - \ft{1}{k(k-1)}\, \int_{\C^2} \alpha_{(0,2)}\wedge {\rm tr}\,{\cal F}^k
\nn\\
&& \int_{\R^4} d^4x\, d^2 \theta\, W(\Phi) =
\int_{\C^2} \alpha_{(2,0)}\wedge {\rm tr}\,  W({\cal F})
\label{topforms}\eea

\subsection{Correlation functions via Localization}

Here we apply localization techniques to the computation of
condensates in ${\cal N}=1$ SYM with various matter content.
Correlators are defined as
\bea
\langle \,{\cal O}\, \rangle_{\cN=1} &=&{1\over V {\cal Z}} \,
\int_{{\cal M}\times \C^2}\,{\cal O} \, {\rm exp}\left[ - S_{{\cal N}=2}
-\int_{\C^2} \alpha_{(2,0)}\wedge {\rm tr}\,  W({\cal F})\right]\label{cor}
\eea
where ${\cal O}$ is any equivariant operator, ${\cal Z}=\langle \,\uno \,\,\rangle$ the
partition function and $V$ the volume.
As we said earlier, to apply the localization theorem ${\cal M}$ must be smooth
and, as a consequence, $\C^2$ must be taken non-commutative.
Let $V$ denotes the volume of the non-commutative space-time
 \be
 V=\int_{\C^2}\, \alpha_{(2,2)}= \bar{z}_0^1  \bar{z}_0^2 z_0^1 z_0^2=\zeta^2
\label{reg}
\ee
This integral is performed using the localization formula which instructs us
to compute at the critical points the zero form part of the equivariant form
to be integrated.
$\zeta$ is the non-commutative parameter\footnote{Given our choice below we have
$\bar z_{1,2}=\zeta \,\partial/\partial z_{1,2}$. While the notion of critical point for a commutative
space is clearly well defined, when a coordinate becomes an infinite dimensional operator this
notion becomes slippery. In all of this work we never need to deal with these type of problems though, and (\ref{reg}) is just taken as a suitable regularization.}
\be
[\bar{z}^j,z^i]=\zeta \,\delta^{ij}
\label{comm}
\ee
The integrals over ${\cal M}$ can be written as a sum over $N$ Young tableaux
$Y$ labelling the fixed points of $\xi$ on the instanton moduli
space \cite{Losev:2003py,Nekrasov:2003rj,Flume:2004rp}
\bea
\langle {\cal O}  \rangle &=&
{1 \over {\cal Z}}  \, \sum_Y \,  {\cal O}_Y \,
 {e^{S_Y}\over {\rm det}^{1\over 2}{\cal L}_{\xi_Y}}\, q^{|Y|}\nn\\
{\cal Z}  &=& \sum_Y \,
 {e^{S_Y}\over {\rm det}^{1\over 2}{\cal L}_{\xi_Y}}\, q^{|Y|}
\label{0p}
\eea
 where $S_Y$  and ${\cal O}_Y$ are the zero form parts of the action $S$ and of the
operator ${\cal O}$.

If  ${\cal O}$ is taken to be $\alpha_{(2,2)}\wedge{\rm tr}\, {\cal F}^J$, since
$S_Y=-{i \over \hbar} \,z_0^1\, z_0^2\, W(\varphi)\to 0$ at the fixed points, we see that
the superpotential term can be always discarded and therefore the
${\cal N}=1$ amplitude reduces to a scalar correlator in ${\cal N}=2$
\be
\langle {\rm tr}\, \varphi^J\, \rangle_{{\cal N}=1}
=\langle {\rm tr}\, \varphi^J e^{-\int_{\C^2} \alpha_{(2,0)}\wedge {\rm tr}\,  W({\cal F})}
\, \rangle_{{\cal N}=2}=\langle {\rm tr}\, \varphi^J\, \rangle_{{\cal N}=2}
\label{n21c}
\ee
To obtain a {\it bona fide} ${\cal N}=1$ correlator we then have to fix the value of the v.e.v.'s
by minimizing the quantum potential as we will discuss in the next section.

If  ${\cal O}$ is taken to be $\alpha_{(0,2)}\wedge {\rm tr}\, {\cal F}^{J+2}$, a power of
$S_Y$ is needed in order to compensate for the unbalanced
$\alpha_{(2,0)}$ form in ${\cal O}$. After this power is extracted
the superpotential term can be discarded leaving again an ${\cal N}=2$
amplitude to compute
\be
\langle {\rm tr}\, \lambda^\alpha \lambda_\alpha \,
\varphi^J\, \rangle_{{\cal N}=1}
=- \frac{1}{c_J V}  \langle  \int \alpha_{(0,2)}\wedge {\rm tr} \F^{J+2}\,
\rangle_{{\cal N}=1}
= \frac{1}{c_J V} \frac{\partial^2}{\partial \tau_w\partial \tau_J}
 {\rm ln} {\cal Z}(\tau_w,\tau_J)|_{\tau_w=\tau_J=0}
\label{n211c}
\ee
with $c_J=(J+2)(J+1)$ and
\be
 {\cal Z}(\tau_w,\tau_J) \equiv \int_\cM {\rm exp}
 \left[-S_{\cN=2} - \tau_w \int_{\C^2} \alpha_{(2,0)}\wedge W(\F)
       - \tau_J \int_{\C^2} \alpha_{(0,2)}\wedge \tr \F^{J+2} \right]
 \label{gen}
 \ee
As for (\ref{n21c}), also this amplitude has to be evaluated at a minimum of the
quantum superpotential.
In writing the gaugino correlator in the form (\ref{n211c})
we use the fact that in the limit
$\hbar\to 0$ only the term proportional to
$\alpha_{(2,0)}\wedge \alpha_{(0,2)}$ contributes
since for unbalanced powers of $\alpha_{(2,0)}$ there is no way to properly contract
the fermionic zero modes.
The equations (\ref{n21c}), (\ref{n211c}) are the generalisation of
the result of \cite{Witten:1994ev}, where the mass deformation
of ${\cal N}=2$ theory has been studied.

Let us now compute the scalar correlators
\bea
\langle {\rm tr}\, \varphi^J  \rangle &=&
{1\over V\,{\cal Z}}\int_{{\cal M}\times \C^2}\,  \alpha_{(2,2)}\wedge
 {\rm tr} \,{\cal F}^J   =
{\hbar^2  \over {\cal Z}}  \, \sum_Y \,  {\rm tr} \varphi_Y^J\,
 {e^{S_Y}\over {\rm det}^{1\over 2}{\cal L}_{\xi_Y}}\, q^{|Y|}\nn\\
  &=& {1\over {\cal Z}} \,\sum_Y \, Z_Y\,
\,{\rm tr}\,\varphi_Y^J\, q^{|Y|}
\label{e-p}
\eea
with $q^{|Y|}=e^{-{8\pi k\over g^2}}$ and
\bea
{\cal Z}_Y &=&\langle\,\, \uno\,\,\rangle={\hbar^2\over {\rm det}^{1\over 2}{\cal L}_{\xi_Y}}\nn\\
{\rm tr}\,\varphi_Y^J &=&{\rm tr}\,\varphi_{\rm cl}^J+\sum_{\alpha=1}^N
\sum_{j_\alpha=1}^{k_{1\alpha}}\left[ (a_\alpha+\hbar(k'_{j_\alpha}-j_\alpha+1))^J
-(a_\alpha-\hbar(j_\alpha-1))^J \right. \nn\\
&&\left.~~~~~~ -(a_\alpha+\hbar(k'_{j_\alpha}-j_\alpha))^J+
(a_\alpha-\hbar \, j_\alpha)^J\right]\label{phij}
\eea
Here $a_\alpha$ are the v.e.v's, $j_\alpha$ runs over the rows of the $Y_\alpha$
tableau, $k'_{j_\alpha}$ is the number of boxes in the column $j_\alpha$
and $k_{1 \alpha}$ are the number of rows in $Y_\alpha$.

The case $J=2$ is particularly simple: (\ref{phij}) depends only on the
number of boxes, $k$, (which is the instanton winding number) in the Young tableaux
\be
{\rm tr}\,\varphi_Y^2={\rm tr}\,\varphi_{\rm cl}^2+2\, k\, \hbar^2
\ee
 Plugging into (\ref{e-p}) one finds the Matone's relation \cite{Flume:2004rp}
 \bea
\langle \ft12\,{\rm tr}\, \varphi^2  \rangle &=& {1\over {\cal Z}} \,\sum_k \, Z_k\,q^k
\,\left[    \ft12{\rm tr}\,\varphi_{\rm cl}^2+ k\, \hbar^2         \right]
=\ft12\,{\rm tr}\,\varphi_{\rm cl}^2+ {\hbar^2 \over {\cal Z}} \,\sum_k \, k\,Z_k\,q^k\nn\\
&=&
\ft12\,{\rm tr}\,\varphi_{\rm cl}^2- q\, \partial_q \, {\cal F}(q)
\label{phi2}
\eea
with
\be
{\cal Z}(q)\equiv\sum_k Z_k\, q^k ; \quad\quad {\cal F}(q)\equiv -\hbar^2 \ln \, {\cal Z}(q)
\ee
 In a similar way one can compute higher point functions. Again correlators
 involving ${\rm tr}\, \varphi^2$ are particularly simple.
   For example
 \bea
\langle {\rm tr}\, \varphi^2 \, {\rm tr}\, \varphi^J\,  \rangle &=&{1 \over {\cal Z}} \,
\sum_Y \, ({\rm tr}\, \varphi_{\rm cl}^2+ 2\,k\, \hbar^2) \,\varphi_Y^J  \, Z_Y\,q^k
=   {\rm tr}\, \varphi_{\rm cl}^2 \, \langle  {\rm tr}\, \varphi^J
\,\rangle\, +{2\, \hbar^2\over {\cal Z}}  \,q\, \partial_q\,\sum_Y\,
 \varphi_Y^J  \, Z_Y\,q^k
\nn\\
&=& {\rm tr}\, \varphi_{\rm cl}^2 \, \langle  {\rm tr}\, \varphi^J
\,\rangle\, +2\, \hbar^2 \,q\, \partial_q \,
\langle {\rm tr}\, \varphi^J \rangle +2\, \hbar^2 \,{{\cal Z}'(q)\over {\cal Z}}
\langle  {\rm tr}\, \varphi^J\rangle\nn\\
&=& \langle {\rm tr}\, \varphi^2 \,\rangle\, \langle  {\rm tr}\, \varphi^J
\,\rangle\, +2\, \hbar^2 \,q\, \partial_q \,
\langle {\rm tr}\, \varphi^J \rangle
\label{fact}
\eea
Notice that in the limit $\hbar\to 0$ the correlator factorizes as expected.
Curiously it is the second term which is relevant to the computations of the
gaugino condensate below.

 Now let us consider correlators involving the gaugino condensate.
 For concreteness we choose
 the superpotential $W(\Phi)=\ft12\, m\, \Phi^2$.
 Applying the localization formula to (\ref{gen}) one finds
 \bea
{1\over V} \frac{\partial^2}{\partial \tau_w\partial \tau_J}
 {\rm ln} {\cal Z}(\tau_w,\tau_J)|_{\tau_w=\tau_J=0}
&=& {m \over \hbar^2} \,\left[
 \langle {\rm tr}\, \varphi^2 \,{\rm tr}\, \varphi^{J+2}\,\rangle
 -\langle {\rm tr}\, \varphi^2\,\rangle\,
 \langle {\rm tr}\, \varphi^{J+2}\,\rangle
 \right]\nn\\
 &=& 2m \,q\, \partial_q \,
 \langle {\rm tr}\, \varphi^{J+2} \rangle
  \label{locgau}
 \eea
 Now comparing (\ref{n211c}) with (\ref{locgau}) we get
 \be
 \langle \tr \lambda\lambda\vphi^J \rangle_{\cN=1} = - \ft{2m}{(J+2)(J+1)}
  \,q\, \partial_q \,
 \langle {\rm tr}\, \varphi^{J+2} \rangle
 \label{final}
 \ee
For $J=0$ one finds
\be
 \langle\, \tr \lambda \lambda \,\rangle = -
{m}\, q\, \partial_q \,\langle {\tr} \varphi^2\rangle
\label{gaugi-cond}
\ee

\section{Pure ${\cal N}=2$ SYM down to ${\cN=1}$}
\setcounter{equation}{0}
Very recently the effective superpotential of ${\cal N}=1$ gauge theories
obtained via deformations of theories with extended supersymmetry has been
computed with matrix model type arguments \cite{Dijkgraaf:2002dh}.
Following this work a general strategy, based on the Konishi anomaly,
has been devised to carry out these computations
in various phases of the
${\cal N}=1$ theory \cite{Cachazo:2002ry}.
See Ref.\cite{Argurio:2003ym} for a nice review. The results
are summarize by the so called ``chiral
ring relations''. The elements of the chiral ring are
those gauge invariant operators that are annihilated
by the $\bar{\cal Q}_{\dot\alpha}$ supersymmetric charges. The notable
properties of this set of operators
are at the heart of supersymmetric instanton calculus and are
responsible for this class of operators to be
independent from space-time variables \cite{Konishi:1989em}.
In \cite{Flume:2004rp}, it was already shown by a direct
multi-instanton computation that ${\cal N}=2$ correlators
of the type $\langle \tr\varphi^J\rangle$ satisfy the
chiral ring relations predicted by Konishi anomaly arguments.
 In this section we extend this result to
scalar and gaugino condensate correlators in the confining
phase of ${\cal N}=1$ SYM.
In these theories gaugino condensates are evaluated via
equations of the type of (\ref{final}).
Following \cite{Cachazo:2002ry} we then introduce generating functions
(resolvent) ${\cal T}(z)$, ${\cal R}(z)$ for the correlators
$\langle \,{\rm tr}\,\varphi^J \rangle$ and
$\langle \,{\rm tr}\,\lambda\lambda \varphi^J \rangle$ respectively
\bea
{\cal T}(z)&=&
\int d^4x\, {\rm tr}\, {1\over z-\varphi}
={1\over z}\sum_k \,\int \alpha_{(2,2)}\wedge  {\rm tr} \,
\left({ {\cal F}\over z}
\right)^k\label{Tz}\\
{\cal R}(z)&=&
\int d^4x\, {\rm tr}\,\lambda \lambda
{1\over z-\varphi}=-{1\over z}\sum_k \,\ft{1}{(k+2)(k+1)}\,\int \alpha_{(0,2)}\wedge
\left({ {\cal F}\over z}
\right)^{k+2}
\label{Rz}
\eea
 The form of $ \langle \,{\cal R}(z)\, \rangle$,
$ \langle \,{\cal F}(z)\, \rangle$ is highly constrained by
the cancellation of Konishi anomaly. Indeed they  are
determined up to two arbitrary polynomials of order $n-1$, $n+1$,
via
\bea
\langle \,{\cal R}(z)\, \rangle &=& W'(z)-\sqrt{W'(z)^2+f(z)}\nn\\
\langle \,{\cal T}(z)\, \rangle &=&  -\ft14\,{c(z)\over \sqrt{W'(z)^2+f(z)}}
\label{ring}
\eea
 The relation (\ref{ring}) is understood in the $\hbar\to 0$ limit.
$W(z)$ is the superpotential and
$f(z),c(z)$ are arbitrary polynomials of order $n-1$, $n+1$ given by
\be
W(z)=\sum_{k=1}^{n+1} \ft1k \,g_k z^k \quad
f(z)=\sum_{k=0}^{n-1} f_k z^k \quad c(z)=\sum_{k=0}^{n-1} c_k z^k
\ee
Different phases of the theory correspond to different choices of the
constants $c_k,f_k$ and account
for partial or total breaking of the gauge group. More precisely
for a generic choice of the $2n$ parameters $c_k,f_k$, the gauge group
is broken down to $\prod_{i=1}^n U(N_i)$ with $c_k,f_k$ parameterizing
the $n$ ranks $N_i$ and the $n$ gaugino condensates $S_i$.
 Here we consider the two extreme cases where the gauge group is completely
broken to $U(1)^N$ or completely unbroken.

\subsection{\bf $U(1)^N$ completely broken phase}

 To break completely the gauge group the superpotential should have $N$
different minima and therefore $W(z)$ should be a polynomial of
order $N+1$. We write
\be W'(z)=P_N(z)=\prod_{\alpha=1}^N\,
(z-a_\alpha) \label{prep}
\ee with $a_\alpha$ parameterizing the
couplings in $W(z)$. The functions $f(z),c(z)$ appearing in
(\ref{n1res}) are given in terms of an associated Seiberg-Witten
like curve \bea
y^2 &=& W'(z)^2+f(z) =  P_N(z)^2-4\,\Lambda^{2 N}\nn\\
c(z) &=& P_N'(z)\label{u1n}
\eea
i.e. $f(z)=-4 \Lambda^{2 N}$. Plugging in (\ref{u1n}) into the chiral
ring relations one finds
\be
 \langle {\cal T}(z) \rangle = {P'(z)\over \sqrt{P(z)^2-4\,\Lambda^{2N}}}
\label{ringt}
\ee
 This is precisely the form of the chiral ring relations for
a pure ${\cal N}=2$ theory with gauge group $U(N)$ and vev's $a_\alpha$.
 In \cite{Flume:2004rp}
 these relations have been shown to hold for ${\cal N}=2$ SYM
with gauge group $SU(2)$ by explicit evaluation of the
multi-instanton determinants. According to (\ref{n21c}) these
results imply that the same chiral ring relations hold for
the ${\cal N}=1$ as well .

\subsection{ $U(N)$ unbroken phase}
 Now we consider the
case where the gauge group is completely unbroken.
For concreteness we choose $W(z)=\ft12\, m\, z^2$, i.e. a mass
deformation of pure ${\cal N}=2$ SYM
\be
W(z)=\ft12 m z^2\quad\quad f(z)=f_0 \quad\quad c(z)=c_0
\label{defn2}
\ee
At the leading order ${1\over z}$ one finds
\bea
S &=& {f_0\over 2 \,m }\nn\\
N &=& -{c_0\over 4 m }
\eea
i.e. the coefficients $f_0=2 m S$ and $c_0= -4 m N$
parameterize the gaugino condensate and the rank $N$ of the gauge group
respectively
($N$ should not to be confused with $n$, the order of $W(z)$).
 Plugging into (\ref{ring}) one finds
\bea
\langle {\cal R}(z) \rangle &=& m z\left(1-\sqrt{1+{2  S\over m z^2}}\right)\nn\\
\langle {\cal T}(z) \rangle &=& { N\over
z\,\sqrt{1+{2  S\over m z^2 }}}\label{n1res}
\eea
Expanding in ${1\over z}$, equations (\ref{n1res}) determine all the
correlators in the chiral ring in terms of the gaugino condensate
$S$ and the rank $N$
of the gauge group.

Notice that for $N=1$, (\ref{n1res}) is of the form
(\ref{ringt}) with $P(z)=z$, $\Lambda^2=- {S\over 2 m}$.
 In general, (\ref{n1res}) can be always put in the ``${\cal N}=2$ form''
(\ref{ringt}) by choosing
\be
 P(z)=\prod_{\alpha=1}^N\, (z-a_\alpha);\quad\quad
 a_\alpha=2\,\Lambda \, \cos \pi {(j-\ft12)\over N} \quad j=1,..N\label{vev}
   \ee
This follows from the identity
\be
 {P'(z)\over \sqrt{P(z)^2-4\,\Lambda^{2N}}}={ N\over
z\,\sqrt{1-{4\Lambda^2\over  z^2 }}}
\ee
satisfied by (\ref{vev}). Therefore the chiral ring relations
for ${\cal N}=1$ SYM theory in the unbroken phase coincide with
those of a pure ${\cal N}=2$ gauge theory with vev's $a_\alpha$
given by (\ref{vev}).
The values $a_\alpha$ are at the minima of the quantum
potential \cite{Douglas:1995nw}.
They represent those points of the moduli space of vacua of the mass perturbed ${\cal N}=2$
theory at which the monopoles are massless. From the geometrical
point of view they are those
degenerate points
at which the elliptical Seiberg-Witten curve has two zeroes of the first order
and all the other ones are of order two.

 Actually there are $N$ different
possible choices for the v.e.v.'s in (\ref{vev}), which are obtained
sending
\be
\Lambda \to \Lambda {\rm e}^{i\pi (\tau +n)/N} \ \ n=0,\ldots, N-1
\label{vacua}
\ee
each choice corresponding to a different vacuum of the ${\cal N}=1$ theory.
All of this is very much in line with the computational strategy in \cite{Shifman:1997nw}.

Finally it is interesting to notice that the chiral ring
observables $\langle \tr \vphi^J\rangle$ for the mass
deformed ${\cal N}=2$ SYM with gauge group $U(N)$
are $N$ times the result for the $U(1)$ case.

\subsection{Multi-instanton tests}

{\bf U(1) case}

We start by considering ${\cal N}=1$ SYM with gauge group $U(1)$ and an adjoint chiral multiplet
with mass $m$. By $U(1)$ instantons we refer to instanton in the non-commutative
theory, but the non-commutative parameter will never enter in our formulae. This
non-commutative extension is also needed in order to compactify the moduli space
of $U(N)$ instantons and will be always understood. The result for the commutative
theory follow from the limit where the commutative parameter is formally turned off.

According to (\ref{n21c}) and (\ref{n211c}) scalar
correlators and gaugino condensates in this theory are related to those in
pure ${\cal N}=2$ SYM.
Scalar correlators in pure ${\cal N}=2$ SYM are given by
(\ref{e-p}) and (\ref{phij}). The instanton partition function ${\cal Z}_Y$
for pure ${\cal N}=2$ is given by
 \be
 Z_Y(\hbar) = \prod_{s\in Y_k} \frac{1}{\hbar^2 L(s)^2} \label{zk}
\ee
 Evaluating (\ref{zk}) one finds
\bea
\langle {\rm tr}\, \vphi^2 \rangle &=& 2 \Lambda^2\nn\\
\langle {\rm tr}\, \vphi^4 \rangle  &=& 6 \Lambda^4\nn\\
\langle {\rm tr}\, \vphi^6 \rangle &=& 20 \Lambda^6\nn\\
\langle {\rm tr}\, \vphi^8 \rangle  &=& 70 \Lambda^8
\eea
in agreement with the chiral ring predictions:
\be
\langle \tr \left( \frac{1}{z  -
\varphi}\right)\rangle = \frac{1}{z}
\frac{1}{\sqrt{1-\frac{4\Lambda^2}{z^2}}}=
\sum_{k=0}^{\infty} \frac{(2k)!}{k!^2}
\frac{\Lambda^{2k}}{z^{2k+1}}
\label{chir-u1}
\ee
i.e.
\be
\langle\tr\varphi^{2k}\rangle= \Lambda^{2k}
\frac{(2k)!}{k!^2} \label{phi-u1}
\ee

{\bf SU(2) case}

The chiral ring relation for scalar correlators in pure ${\cal N}=2$ SYM
with gauge group $SU(2)$ were tested in \cite{Flume:2002az} against multi-instanton
calculus. Here we recall these results and extract from them the scalar correlators
for the mass deformed ${\cal N}=1$ theory.
 The Seiberg-Witten $SU(2)$ curve is given by the equation
\be
y^2 =  P(z)^2-4\,\Lambda^4=(z^2-\frac{u}{2})^2-4\,\Lambda^4
\label{curSW}
\ee
so that expanding the resolvent
\be
 \langle\tr  \frac{1}{z-\vphi} \rangle = {P'(z)\over \sqrt{P(z)^2-4\,\Lambda^4}}
\label{ringt1}
\ee
we find $u=<\tr\vphi^2>$.
The minima of the quantum potential associated to the $SU(2)$ unbroken phases,
i.e. the points where two zeros of the Seiberg--Witten curve
are colliding, are given by $u_{1,2} =\pm 4\Lambda^2$ that correspond to
$u=\pm(a_1^2+a_2^2)$ with $a_\alpha$ given by (\ref{vev}) for $N=2$.

 By a direct computation in the ${\cal N}=2$ SYM one finds \cite{Flume:2002az}
\bea
u= \langle {\rm tr}\, \varphi^{2}
\rangle &=& 2 \,a^2+{q\over a^2}+\ft{5}{16}\, {q^2\over a^6}+
\ft{9}{32}\, {q^3\over a^{10}}+\ldots\nn\\
\langle {\rm tr}\, \varphi^{4}
\rangle &=& 2 \,a^4+6 \, q+\ft{9}{8}\, {q^2\over a^4}+
\ft{7}{8}\, {q^3\over a^{8}}+\ldots\nn\\
\langle {\rm tr}\, \varphi^{6}
\rangle &=& 2 \,a^6+15 \, q\, a^2+\ft{135}{16}\, {q^2\over a^2}+
\ft{125}{32}\, {q^3\over a^{6}}+\ldots
\label{eqnvecchiolavoro}
\eea
The $a$ here should not be confused with its quantum analog $a_\alpha$
entering in the Seiberg-Witten curve and given by (\ref{vev}).
The right variable for a comparison is not the classical v.e.v. $a$ but
the ``quantum" coordinate $u$.
Therefore, inverting the first of these equations, we get
\beq
a=\sqrt{\frac{u}{2}}-\frac{1}{{\sqrt{2}}\,u^{\frac{3}{2}}}q-\frac{15}
{4\,{\sqrt{2}}\,u^{\frac{7}{2}}}q^2-\frac{105}
{4\,{\sqrt{2}}\,u^{\frac{11}{2}}}q^3+\ldots
\ee
Substituting back in (\ref{eqnvecchiolavoro}) we get\footnote{For the sake of conciseness
in (\ref{eqnvecchiolavoro}) we did not write the gravitational corrections to the results.
They can be found in \cite{Flume:2002az} together with the corrections to the formula below.}
\begin{eqnarray}
\langle \tr \vphi^4\rangle &=&4 q+\frac{u^2}{2} \, , \nonumber \\
\langle \tr \vphi^6\rangle &=&6qu + \frac{u^3}{4} \, , \nonumber \\
\langle \tr \vphi^8\rangle &=&12 q^2 + 6 q u^2 + \frac{u^4}{8}
\label{su2ringrelations}
\end{eqnarray}
which are exactly the chiral ring relations. We draw the reader's attention on the fact that even if
the expansion parameter in (\ref{eqnvecchiolavoro}) is close to $1$, since we are in the strong coupling region,
still all our manipulations are safe being the relations (\ref{su2ringrelations}) exact.
Plugging $u_{1,2} =\pm 4\Lambda^2$ in (\ref{su2ringrelations})
and comparing with
(\ref{phi-u1}) one easily see that the $SU(2)$ scalar condensates
are twice the $U(1)$ ones, in agreement with the chiral prediction
(\ref{n1res}).
We stress that this procedure is just the extension of the strategy of Ref.\cite{Shifman:1997nw}.

{\bf U(N) case}

According to (\ref{n1res})
$U(N)$ scalar correlators are given by $N$ times the
result for $U(1)$ i.e.
\be
\langle\tr\varphi^{2k}\rangle= N\Lambda^{2k}
\frac{(2k)!}{k!^2} \label{phi-un}
\ee
 These results follows from those of ${\cal N}=2$ choosing
 vev given by (\ref{vev}). For this choice of vev, the chiral
ring relation (\ref{ring}) for ${\cal R}(z)$ becomes
\be
{\cal R}(z) = \langle \tr \left(\frac{\lambda\lambda}
{z-\vphi}\right)\rangle
=m z \left(1-\sqrt{1-\frac{4\Lambda^{2}}{z^2}}\right)
\ee
By comparing the expansion of the above formula with
(\ref{phi-un}) we get the relation
\be
\langle {\rm tr}\lambda \lambda \vphi^{2k}\rangle
 = - \frac{m}{(2k+1)N} \langle {\rm tr} \vphi^{2k+2}\rangle
 = - \frac{m}{(2k+1) (2k+2)} \, \Lambda_N \partial_{\Lambda_N}
 \langle {\rm tr} \vphi^{2k+2}\rangle
\label{R-u1}
\ee
with $\Lambda_N\equiv \Lambda^{1/N}$.
 This result matches with the computation via localization
presented in (\ref{final}) with $J=2k$ and $q=\Lambda^2_N$.

In the limit $m\to\infty$, in which the massive particles decouple,
from (\ref{R-u1}) we can recover the results for pure ${\cal N}=1$ SYM.

\vspace{.5cm}
\section{Chiral ring relations for ${\cal N}=1^*$ SYM}
\setcounter{equation}{0}

  In this section we consider the ${\cal N}=1^*$ gauge theory, defined as a mass
 deformation of ${\cal N}=4$ where all three chiral multiplets have
 mass $m$. Alternatively the ${\cal N}=1^*$ theory can be defined
 as a mass deformation of ${\cal N}=2^*$. According to the
 discussion in Sect.2 scalar correlators and gaugino condensates in ${\cal N}=1^*$
 are related to scalar correlators in ${\cal N}=2^*$ via (\ref{n21c}), (\ref{n211c}).
 As before the ${\cal N}=2^*$ amplitudes should be evaluated at the v.e.v.   specifying
 the ${\cal N}=1$ vacuum, i.e. the points in the moduli space where the
 Seiberg-Witten curve degenerates and the $SU(N)$ symmetry is unbroken.

\subsection{$U(1)$ group}

Here the situation is particularly simple since the v.e.v. of the scalar
field can be shifted to zero by a redefinition since the theory has only one vacuum.
Using the results of \cite{Bruzzo:2002xf} (and setting for simplicity $\epsilon_1=-\epsilon_2= \hbar$)
one can write
\bea
\langle
\tr(\vphi^2)\rangle_{{\cal N}=2^*}&=&-\frac{\hbar^2}{Z(q)}\sum_k k Z_k q^k
\eea
where ${ Z}_k$ is the $k$--instanton partition function given by
\be
Z_k(\hbar) = \sum_{\{Y_k\}}\prod_{s\in Y_k} \Bigl[ 1 -
\frac{m^2}{E(s)^2}\Bigr] \label{zk1}
\ee
Here $E(s)= -\hbar L(s)$, with $L(s)$ the length of the hook centered
in the $s$--box of the Young tableau and stretched between the top and left end
of the tableau. For the first few instanton numbers one finds:
\bea
Z_{\tinyyoung{\hfill}}&=&  1-\frac{m^2}{\hbar^2} \label{z1} \nn\\
2\,Z_{\tinyyoung{\hfill\hfill}}&=& 2\,\left(1 - \frac{m^2}{4\hbar^2}\right)
        \left(1 - \frac{m^2}{\hbar^2}\right)\nn\\
2\,Z_{\tinyyoung{\hfill\hfill\hfill}}&=&2\, \left( 1 -
\frac{m^2}{9\hbar^2}\right) \left( 1 - \frac{m^2}{4\hbar^2}\right)
\left( 1 - \frac{m^2}{\hbar^2}\right) \nn\\
Z_{\tinyyoung{\hfill\hfill,\hfill}}&=&
  \left( 1 - \frac{m^2}{9\hbar^2}\right) \left( 1 -
\frac{m^2}{\hbar^2}\right)^2 \label{zy} \eea
 The factor $2$ in the second and third equations come from the identical contributions of a diagram and its transpose.
 Putting them together we obtain
  \bea
   \langle
\tr \vphi^2 \rangle_{{\cal N}=1^*}&=&   (m^2-\hbar^2)\,(1+q+3 q^2+4 q^3+7 q^4+\ldots)
\nn\\
&=&(m^2-\hbar^2)\,\sum_{d|k} d \,q^k =
-\ft{1}{24}\,(m^2-\hbar^2)\, E_2(q) \label{DVphi2}
\eea
This formula is exact and includes the gravitational corrections: the term multiplied by $\hbar^2$
is the only gravitational correction to this correlator.
Finally, taking the $\hbar\to 0$ limit, applying (\ref{gaugi-cond}) and using the
relation $\partial_\tau E_2(\tau) = \frac{i\pi}{6}(E_2^2 - E_4)$
we get the gaugino condensate \be \langle\, \tr \lambda^\alpha
\lambda_\alpha \,\rangle_{{\cal N}=1^*} = \frac{m^3}{24\cdot
12}(E_2^2(q) - E_4(q)) \label{DVg} \ee

\subsection{Scalar correlators in $SU(2)$ }

The situation in this case is more subtle since some extra work is required to find out the exact value of the v.e.v. in the ${\cal N}=1$ vacua.
The Seiberg-Witten curve for the SYM with gauge group $SU(2)$ is given by a genus one Riemann surface.
Indeed the case in which matter gets massive was analyzed in \cite{Seiberg:1994aj}
with methods similar to those
in \cite{sw}. For gauge groups of higher rank we need some other input
which was provided by the connection
with integrable models \cite{Gorsky:1995zq,Martinec:1995by,Donagi:1995cf} and will be
the subject of one of the next subsections.
Let us start with the so called ${\cal N}=2^*$ theory \cite{Seiberg:1994aj} and its Seiberg-Witten curve
\be
y^2=\prod_{i=1}^3(x-{\cal E}_i) \quad\quad
{\cal E}_i=e_i\frac{\tilde u}{2}+\frac{1}{4}e_i^2m^2   \label{curve}
\ee
Following \cite{Erdelyi}, the $e_i$'s are given by
\bea
e_1&=&\ft13 \left(\frac{\pi}{2\omega_1}\right)^2\left[\vartheta_3^4(0\vert q)+
\vartheta_4^4(0\vert q)\right]\nn\\
e_2&=&\ft13 \left(\frac{\pi}{2\omega_1}\right)^2\left[\vartheta_2^4(0\vert q)-
\vartheta_4^4(0\vert q)\right]\nn\\
e_3&=&-\ft13
\left(\frac{\pi}{2\omega_1}\right)^2\left[\vartheta_3^4(0\vert q)+
\vartheta_2^4(0\vert q)\right]
\label{eis}
\eea
with $q=e^{2\pi i
\tau}$, $\tau=\omega_2/\omega_1$ and $\omega_{1,2}$ the semi-periods
\footnote{Apart from this choice of $q$ we adopt the
notation in \cite{Erdelyi}.}. Notice that
$e_1+e_2+e_3=0$. To simplify the computations in this subsection
we choose the semi-period $\omega_1=\pi/2$. Given this choice, the
$q$-expansion of the $e_i$'s is given in
(\ref{eisexp}) in the Appendix. The variable $\tilde{u}$ is
related to the physical one, $u=\langle\tr \phi^2\rangle$, by \be
\tilde u=u-m^2\left(\frac{1}{6}+\sum_{n=1}^\infty\alpha_n
q^n\right) \ee In \cite{Seiberg:1994aj} a guess is given for the
form of the coefficients $\alpha_n$ which does not seem to agree
with an explicit instanton computation of winding number one
\cite{Dorey:1996ez}. Our first task in this section is to compute
all the $\alpha_n$'s.

The variable $\tilde u$ is related to the vev $a$ in such a way
that ${\partial a\over \partial \tilde{u}}$ and ${\partial a\over
\partial \tilde{u}}$ reproduce the periods of the Seiberg-Witten
curve (\ref{curve}), i.e. via the differential
equations\footnote{The dictionary between the variables employed
here and those in \cite{Dorey:1996ez} is $v_{DKM}=\sqrt{2}a,
u_{DKM}=u/2$. Moreover these authors follow the notations of \cite{Chandra} in which $e_2$ and $e_3$ are
interchanged with respect to our choice. Finally their $\omega_1$ is the period and not the semi-period.}
\be \frac{\partial a}{\partial \tilde u}=
\frac{1}{2\pi}\frac{K(k^2)}{\sqrt{{\cal E}_1-{\cal E}_2}}
~~~~~\quad\quad \frac{\partial a_D}{\partial \tilde u}=
\frac{1}{2\pi}\frac{K(k'^2)}{\sqrt{{\cal E}_1-{\cal E}_2}} \ee
with $k^2=({\cal E}_3-{\cal E}_2)/({\cal E}_1-{\cal E}_2)$ and
$k'^2=({\cal E}_1-{\cal E}_3)/({\cal E}_1-{\cal E}_2)$ and
$K(k^2)$ is the complete elliptic integral of first kind
(\ref{kint}).
 Expanding in $q$ one finds
\bea
&&\frac{\partial a}{\partial \tilde u}=
\ft12 z +
   \ft{1}{8}m^2\, z^3\,
      \left( 4 + 3\,m^2\,z^2 \right) \,
      q^2 + \ft{3}{128}\,m^2\,
      z^3\,
      \left( 64 + 128\,m^2\,z^2 +
        35\,m^6\,z^6 \right) \,q^4
\label{dau}\\
   && +\ft{1}{512}\,m^2\,z^3\,
      \left( 1024 + 4608\,m^2\,z^2 +
        1600\,m^4\,z^4 +
        6720\,m^6\,z^6 -
        1260\,m^8\,z^8 +
        1155\,m^{10}\,z^{10} \right) \,
      q^6 +\ldots \nn
\eea
 with $z=(\ft12\tilde u+\ft{1}{12} m^2)^{-{1\over 2}}$.
In order to
find $\tilde{u}$ we integrate (\ref{dau})
 (fixing the integration constant to zero to satisfy monodromies)
 and invert this equation. One finds
\bea
\tilde u &=&
2 \, {a}^{2}-\frac{{m}^{2}}{6}
+\left ( 4 \, {m}^{2}+\frac{{m}^{4}}{{a}^{2}}\right )  \, q+\left ( 12 \, {m}^{2}
+\frac{6 \, {m}^{4}}{{a}^{2}}-\frac{3 \, {m}^{6}}{{a}^{4}}
+\frac{5 \, {m}^{8}}{16 \, {a}^{6}}\right )  \, {q}^{2}\nn\\
&&
+\left ( 16 \, {m}^{2}+\frac{12 \, {m}^{4}}{{a}^{2}}
-\frac{24 \, {m}^{6}}{{a}^{4}}+\frac{15 \, {m}^{8}}{{a}^{6}}
-\frac{7 \, {m}^{10}}{2 \, {a}^{8}}
+\frac{9 \, {m}^{12}}{32 \, {a}^{10}}\right )  \, {q}^{3}
+\ldots
\label{utilde}
\eea
 This can be compared with the results in \cite{Bruzzo:2002xf,Flume:2004rp}
 for $u=\langle\tr \phi^2\rangle$. In the limit $\hbar\to 0$
\begin{eqnarray}
u&=&\langle\tr \varphi^2\rangle=
2a^2+\big(-4m^2+\frac{m^4}{a^2}\big) q+\big(-12m^2+\frac{6 m^4}{a^2}-\frac{3 m^6}{a^4}+
\frac{5m^8}{16a^6}\big)q^2\nonumber\\
&+&\big(
-16m^2+\frac{12m^4}{a^2}-\frac{24m^6}{a^4}+\frac{15m^8}{a^6}-\frac{7m^{10}}{2a^8}+
\frac{9m^{12}}{32a^{10}}\big)q^3 +\ldots
\label{u_inverted}\end{eqnarray}
Comparing (\ref{utilde}) and (\ref{u_inverted})
 one finds
\be
\tilde{u}=u-\ft16 m^2+ 8 m^2\, \sum_{d|k} d \, q^k=u+\ft16 m^2-\ft13\, m^2\, E_2(q)
\label{alfan}
\ee
 To extract from the ${\cal N}=2^*$ result (\ref{u_inverted})
 the result for the ${\cal N}=1^*$ theory we need to specify
 the value of $\tilde{u}$ which corresponds to the points in the moduli space where two solutions of
the cubic (\ref{curve}) are coincident i.e.
$\tilde u=\frac{m^2}{2}e_i$. Plugging this back into (\ref{alfan})
and using the identities (\ref{e-id}) displayed in the Appendix one finds
\begin{eqnarray}
u_1
&=&\frac{m^2}{6}( 4\, E_2(q^2)-1)\nn \\
u_2
&=&\frac{m^2}{6}(E_2(-\sqrt{q})-1)\nn \\
u_3
&=&\frac{m^2}{6}(E_2(\sqrt{q})-1)\label{uvalues}
\end{eqnarray}
$u_1$ gives the result in
the Higgs phase, while in the confining phase $u=\ft12(u_{2}-u_{3})$,  
with $u_{2,3}$ the contributions coming from the two vacua.
The results for pure ${\cal N}=1$ gauge theory can be read from this
by sending $m\to \infty$ keeping $\Lambda^2_{{\cal N}=1}=m^2 q^{1\over 2}$
in the confining phase:
$u_{2,3}=\pm 4\Lambda^2_{{\cal N}=1}$.

In a similar way one can consider higher scalar correlators.
Once again following \cite{Bruzzo:2002xf,Flume:2004rp} we find
\bea
&&\langle\tr\varphi^4\rangle =
2\,a^4 - 6\,m^2\,\left( 4\,a^2 - m^2 \right) \,q +
  \left( -72\,a^2\,m^2 + 48\,m^4 -
    \frac{18\,m^6}{a^2} + \frac{9\,m^8}{8\,a^4}
    \right) \,q^2\nonumber\\
    &&~~~~~+
\left( -96\,a^2\,m^2 + 152\,m^4 -
    \frac{156\,m^6}{a^2} + \frac{72\,m^8}{a^4} -
    \frac{55\,m^{10}}{4\,a^6} +
    \frac{7\,m^{12}}{8\,a^8} \right) \,q^3
+\ldots
\label{fi4}
\eea
Inverting (\ref{u_inverted}) we find
\bea
a&=&\sqrt{\frac{u}{2}}-q\,\left( \frac{m^4}{{\sqrt{2}}\,u^{\frac{3}{2}}} -
    \frac{{\sqrt{2}}\,m^2}{{\sqrt{u}}} \right)+
q^2\,\left( \frac{-15\,m^8}
     {4\,{\sqrt{2}}\,u^{\frac{7}{2}}} +
    \frac{6\,{\sqrt{2}}\,m^6}{u^{\frac{5}{2}}} -
    \frac{4\,{\sqrt{2}}\,m^4}{u^{\frac{3}{2}}} +
    \frac{3\,{\sqrt{2}}\,m^2}{{\sqrt{u}}} \right)\nonumber\\&&+
q^3\,\left( \frac{-105\,m^{12}}
     {4\,{\sqrt{2}}\,u^{\frac{11}{2}}} +
    \frac{245\,m^{10}}{2\,{\sqrt{2}}\,u^{\frac{9}{2}}} -
    \frac{90\,{\sqrt{2}}\,m^8}{u^{\frac{7}{2}}} +
    \frac{53\,{\sqrt{2}}\,m^6}{u^{\frac{5}{2}}} -
    \frac{12\,{\sqrt{2}}\,m^4}{u^{\frac{3}{2}}} +
    \frac{4\,{\sqrt{2}}\,m^2}{{\sqrt{u}}} \right)
+\ldots
\nn\\
\eea
Finally plugging this inside (\ref{fi4}) we get
\be
\langle\tr\varphi^4\rangle=\frac{u^2}{2}-8m^2(q+3q^2+4q^3+\ldots)u-4m^4(-q+q^2+28q^3+\ldots)
\label{chiralringtype}
\ee
 In the next section we will express the terms multiplied by $m^2$ and $m^4$ in (\ref{chiralringtype})
 in terms of known functions using the chiral ring relations of ${\cal N}=2^*$ SYM.

\subsection{Chiral ring relations}

In this section we will derive the chiral ring relations for the ${\cal N}=1^*$
scalar correlators and test them against the multi-instanton results in
the previous section. Here we focus on the $SU(2)$ case. Higher rank
groups will be studied in next subsection exploiting the connection between
SYM and integrable models.
Following \cite{Nekrasov:2003rj} we define the following generating
function for correlators with arbitrary powers of the scalar fields\footnote{
In our notations $\varphi_{\rm cl}={\rm diag}\{a,-a\}$ and $\tr {\bf 1}=2$.}
\bea
G(z)&=&
\Big\langle\tr \frac{1}{z-\vphi- \ft{i}{2} {\mmm}} \Big\rangle -
\Big\langle\tr  \frac{1}{z-\vphi+\ft{i}{2} {\mmm}} \Big\rangle
\label{resolventwithg} \eea $G(z)$ is an analytic function with
branch cuts in $[\alpha_i^-\pm \ft{i}{2} {\mmm},
\alpha_i^+\pm \ft{i}{2} {\mmm}],\,\,i=1,\ldots,N$. Let $U$ be its
domain of definition\footnote{To compare with
\cite{Nekrasov:2003rj} we should replace $m$ by its analytic
continuation $i{\mmm}$.}. Out of $G(z)$ we can build the function
\cite{Kazakov:1998ji} \be \varpi(z)=\frac{1}{2\pi i}\int_\infty^z
G(y)dy \label{gzint} \ee a map from $U$ to the cover, ${\cal
C}_N$, of an elliptic curve, ${\cal C}$. ${\cal C}$ is obtained
identifying the points $(\varpi+1,z)\sim (\varpi,z)$,
$(\varpi+\tau,z)\sim (\varpi,z+i{\mmm})$. In \cite{D'Hoker:1997ha}
a function $f(z,\varpi)$ with this double-periodic property was
introduced \bea
f(z,\varpi)&=&\frac{1}{\vartheta_1(\varpi\vert\tau)}
\sum_{n=0}^N\frac{1}{n!}\frac{\partial^n}{\partial \varpi^n}
\vartheta_1(\varpi\vert\tau)\left(-i{\mmm}\frac{\partial}{\partial z}\right)^nH(z)\nonumber\\
&=&\sum_{n=0}^N \frac{h_n(\varpi)}{n!}
\left(-i{\mmm}\frac{\partial}{\partial z}\right)^nH(z)
\label{spectralcurve} \eea
 with
\be h_n={1\over \vartheta_1(z|\tau)}\, {\partial^n\over \partial
z^n}\, \vartheta_1(z|\tau)\quad\quad~~~~ H(z)=\prod_{i=1}^N
(z-z_i) \ee and $z_i$ free parameters specifying the vacuum. In
the above formulae and in the following computations we choose
$\omega_1=\ft12$.
When comparing with the results of the previous
subsection we will take care of reintroducing the suitable factors
of $\pi$.
It is then natural to define $\varpi(z)$ as the solution
of \be f(-2\pi i z,\varpi)=0 \label{curveeq} \ee For the $SU(2)$
case we have $H(z)=z^2-z_1^2$ and\footnote{This form of the
equation can be immediately compared with that of
\cite{Donagi:1995cf}. In this last reference the equation for
${\cal C}_N$ is given in the form
$P_N+A_2P_{N-2}+\ldots+A_NP_0=0$. The $P_i$ are polynomials in $z,
\wp(\varpi)$. Setting $2\pi it=-2\pi iz-i{\mmm}h_1(\varpi)$ we find
$0=f(2\pi
it,\varpi)=-4\pi^2t^2+(i{\mmm})^2h_1^\prime-z_1^2=-4\pi^2t^2-(i{\mmm})^2\wp(\varpi)-
(i{\mmm})^2\pi^2E_2(q)/3-z_1^2
=-4\pi^2[t^2+x+(i{\mmm})^2E_2(q)/12+z_1^2/4\pi^2]$. The two last
terms in this equation give an explicit expression for the $A_2$
introduced above. In turn, according to \cite{Donagi:1995cf},
$A_2$ should be identified with $-\tilde u/2$ thus giving (we
revert to using the physical mass) $\tilde
u=-m^2E_2(q)/6-z_1^2/2\pi^2$. This relation will also be found
from the chiral ring relations later on.} \be
f(z,\varpi)=z^2-2i{\mmm} z
h_1(\varpi)+(i{\mmm})^2h_2(\varpi)-z_1^2 \label{curva} \ee The
resolvent $G(z)$ then follows from (\ref{gzint}) \be G(z)=2\pi i
\, \varpi(z)^\prime \label{fandg} \ee To find an explicit
expression for (\ref{fandg}) we should first compute $\varpi(z)$
 from (\ref{curveeq}).  We need
\bea
&&h_1(\varpi)
=\frac{\vartheta_1(\varpi\vert\tau)^\prime}{\vartheta_1(\varpi\vert\tau)}
=\pi\cot\pi
\varpi+4\pi\sum_{n=1}^\infty \frac{q^n}{1-q^n}\sin 2\pi n\varpi=\frac{1}{\varpi}
+\left( \frac{-\pi^2}{3}\right.\nonumber\\&&
 \left.  + 8\,f_1(q)\,\pi^2 \right) \,\varpi +
  \left( \frac{-{\pi }^4}{45} - \frac{16\, f_3(q){\pi }^4}{3} \right)
    \varpi^3 + \left( \frac{-2\,{\pi }^6}{945} +
    \frac{16\, f_5(q){\pi }^6}{15} \right)\,\varpi^5+\ldots
    \nn\\
&& h_2(\varpi)=\frac{\vartheta_1(\varpi\vert\tau)^{\prime\prime}}
{\vartheta_1(\varpi\vert\tau)}=
h_1(\varpi)^\prime+h_1(\varpi)^2
\eea
with $f_p(q)$ defined by (\ref{fps}).
 Out of the two roots of (\ref{curveeq}) we choose that one with a
  pole at $\varpi=0$. Expanding around $\varpi=0$ one finds
\bea &&z={1\over 2\pi}\left(-{\mmm} h_1-\sqrt{ -{\mmm}^2\,
h_1'-z_1^2}\right) =-\frac{{\mmm}}{\pi \,\varpi}+
\left[\frac{E^2\,\pi }{{\mmm}} + \left( \frac{1}{12}-2\,f_1(q)
\right) \,{\mmm}\,\pi \right] \,\varpi\nn\\
&&+
\left[\frac{E^2\pi^2}{6}\left(-1+24\,f_1(q)\right)+\frac{E^4\pi^2}{{\mmm}^2}+
\frac{{\mmm}^2\pi^2}{720}\left(1-240\,f_1(q) + 2880\,{f_1(q)}^2-
960\,f_3(q)
\right)\right]\frac{\pi\varpi^3}{{\mmm}}\nn\\
&&+ \left[E^2(\ft{1}{120}-2\,f_1(q) +24\,f_1(q)^2-8\,f_3(q))-
{E^4\over
{\mmm}^2}\frac{1}{2}+12\,f_1(q))+\frac{2E^6}{{\mmm}^4}+\frac{{\mmm}^2}{30240}
\right.\label{seriestoinvert}\\&&\left. +
{\mmm}^2\left(\frac{f_1(q)\,}{60} -2\,{f_1(q)}^2\, +16f_1(q)^3\,
+\frac{2\,f_3(q)\,  }{3} -16\,f_1(q)\,f_3(q)\, +\frac{4\,f_5(q)\,
}{5}\right)\right]\frac{\pi^5\varpi^5}{{\mmm}}+\ldots \nn \eea
where $E^2=z_1^2/(4\pi^2)$. Inverting (\ref{seriestoinvert}) we
find \bea &&\varpi(z)=-\frac{{\mmm}}{z\,\pi}
-\frac{{\mmm}^3}{12\pi z^3}\left( -1 + 24\,f_1(q) -
\frac{12\,E^2}{{\mmm}^2}\right)-\frac{1}{\pi z^5}\left[{\mmm} E^4
-\frac{E^2}{2}\,\left( -1 + 24\,f_1(q) \right) \,{\mmm}^3
\right.\nonumber\\&&\left.
  + \left( 3 - 80\,f_1(q) + 960\,f_1(q)^2+ 320\,f_3(q)\right)\frac{{\mmm}^5}{240}\right] + \ldots
\label{seriesinverted} \eea Taking the derivative and comparing
the result for $G(z)=2 \pi i \varpi'(z)$ with the expansion of
(\ref{resolventwithg}) \be
G(z)=\frac{2i{\mmm}}{z^2}+\frac{1}{z^4}\left(\frac{(i{\mmm})^3}{2}+
3i{\mmm}\langle\tr\varphi^2\rangle\right)+
\frac{1}{z^6}\left(\frac{(i{\mmm})^5}{8}+\frac{5}{2}(i{\mmm})^3\langle\tr\varphi^2\rangle+
5i{\mmm}\langle\tr\varphi^4\rangle\right) +\ldots
\label{functiongg} \ee one finds \bea \langle\tr\varphi^2\rangle
&=& 4\,f_1(q)\,{\mmm}^2-\frac{z^2_1}{2\pi^2}
\label{k1_vs_u}\\
\langle\tr\varphi^4\rangle
&=&\frac{(\langle\tr\varphi^2\rangle)^2}{2}
+8{\mmm}^2f_1(q)\langle\tr\varphi^2\rangle+{\mmm}^4\left(\frac{4}{3}f_1(q)-32f_1(q)^2+\frac{8}{3}f_3(q)\right)
\label{ring-n2*} \eea and so on. Formula (\ref{ring-n2*}) agrees
with (\ref{chiralringtype})  (once we take $i{\mmm}\to m$) and
confirms the multi-instanton result. Formula (\ref{k1_vs_u})
relates $z_1$ to $u$. Substituting (\ref{alfan}) in
(\ref{k1_vs_u}) we find
\begin{equation}
\frac{z_1^2}{2 \pi^2}=-\tilde{u}-\frac{{\mmm}^2}{6}+4 {\mmm}^2
f_1(q),
\end{equation}
which is the result obtained in the footnote of the previous page
from the definition of the Seiberg-Witten curve.
In the remaining of this section we compute
the $z_1$ corresponding to the $SU(2)$ ${\cal N}=1^*$
vacua.

The $SU(2)$ unbroken vacuum is given by choosing $z_1$ such that
the corresponding Seiberg-Witten curve degenerates. Solving for
$z$ in (\ref{curva}), we find
\begin{equation}
2\pi  z={\mmm}\left(-h_1(\varpi)\pm
\sqrt{-\frac{z_1^2}{{\mmm}^2}-h_1^\prime(\varpi)}\,\right)={\mmm}\left(-h_1(\varpi)\pm
\sqrt{-\frac{z_1^2}{{\mmm}^2}+\frac{\pi^2}{3}E_2(q)+\wp(\varpi)}\,\right)
\end{equation}
The positions of the end points of the
cuts are defined by\footnote{The cuts are the regions where there are the allowed values of the spectrum.
At the end points of this regions the velocity must be zero.}
\begin{equation}
2\pi \frac{dz}{d\varpi}={\mmm}\left(-h_1^\prime(\varpi)\pm
\frac{\wp^\prime(\varpi)}{2\sqrt{-\frac{z_1^2}{{\mmm}^2}+\frac{\pi^2}{3}E_2(q)
+\wp(\varpi)}}\,\right)=0
\end{equation}
or
\begin{equation}
\sqrt{-\frac{z_1^2}{{\mmm}^2}+\frac{\pi^2}{3}E_2(q)
+\wp(\varpi)}=\mp \frac{\wp^\prime(\varpi)}{2 h_1^\prime(\varpi)}.
\end{equation}
The condition that the branch points collide should imply that the
square root vanishes which according to the last equality is
equivalent to
\begin{equation}
\wp^\prime(\varpi)=0 \label{pprime}
\end{equation}
$\varpi=\omega$, $\omega^\prime$ and $\omega+\omega^\prime$ are
the zeroes of $\wp^\prime(\varpi)$ and at these points
$\wp(\varpi)$ takes the values
$\pi^2 e_1$, $\pi^2 e_3$ and $\pi^2 e_2$ respectively\footnote{Since we compare with a result in the
previous subsection, we must take care of the difference in our two choices of semi-period
($\omega_1=\pi/2$ then and $\omega_1=1/2$ now. See the appendix for more details}.
Thus at degenerated points we have
\begin{equation}
{z_1^2\over 2 \pi^2}=\frac{{\mmm}^2}{6}E_2(q)+{{\mmm}^2\over 2}\,
e_i, \quad\quad i=1,2,3. \label{z1s}
\end{equation}
Plugging this into (\ref{k1_vs_u}) we get \be
u_i=\langle\tr\varphi^2\rangle = \frac{{\mmm}^2}{6}-
\frac{{\mmm}^2}{3} E_2(q)-\frac{{\mmm}^2}{2} e_i \ee in agreement
with (\ref{alfan}) after replacing $i{\mmm}\to m$ and using
$\tilde u_i=\frac{m^2}{2} e_i$

\subsection{$SU(N)$ case and Integrable models}

The general $SU(N)$ case can be studied exploiting the description
of ${\cal N}=1^*$ in terms of the Calogero-Moser integrable
system. This system is defined by the Hamiltonian \be
H=\ft12\sum_{i=1}^N  p_i^2+\ft{1}{2}\,m^2\,\sum_{i<j}^N{\wp}
(x_i-x_j) \ee At equilibrium $p_i=0$ and ${\wp}'(x_i-x_j)=0$.
This generalizes the condition (\ref{pprime}) for $SU(2)$. As in
the $SU(2)$ case we can introduce a curve ${\cal C}_N$, an
$N$-fold branch cover of ${\cal C}$, via \bea
0&=&f(z,\varpi)=f(\lambda+mh_1(\varpi),\varpi))=\det(\lambda\,\, \uno_{\,\,N\times N}-L(\varpi))\nn\\
&=&\lambda^N+\sum_{k=1}^NJ_k(\varpi)\lambda^{N-k}=\prod_{k=1}^N(\lambda-\lambda_i)
\label{spectraldet}
\eea
For each value of $\varpi$ in (\ref{spectraldet}), which is the
spectral parameter of ${\cal C}$,   we find
$N$ values of $\lambda$ which describes ${\cal C}_N$.
The matrix $L(\varpi)$ together with a matrix $M(\varpi)$, form a Lax pair for the
Calogero-Moser system with spectral parameter $(\varpi)$ \cite{Krichever:1980}.
Their explicit form is \cite{Olshanetsky:1981dk}
\bea
L_{ij}(\varpi)&=&p_i\delta_{ij}-m(1-\delta_{ij})\Phi(x_i-x_j,\varpi)\nonumber\\
M_{ij}(\varpi)&=&m\delta_{ij}\sum_{k\neq i}\wp(x_i-x_k)-m(1-\delta_{ij})\Phi^\prime(x_i-x_j,\varpi)
\label{laxpair}
\eea
where
\be
\Phi(x,\varpi)=\frac{\sigma(\varpi-x)}{\sigma(\varpi)\sigma(x)}e^{\zeta(\varpi)x}
\label{sigmazeta}
\ee
$\sigma$ is the standard Weierstrass function \cite{Erdelyi}.
The coefficients $J_k$'s in (\ref{spectraldet})
are connected with the integrals of motions
\be
I_k=\frac{1}{k}\tr(L^k)= \frac{1}{k}\left(\sum_{i=1}^N\lambda_i\right)^k
\ee
$I_2$ is the Hamiltonian of the classical system and it is connected with the variable $u$ in
(\ref{u_inverted}) \cite{Dorey:1999sj,Dorey:2002ad}.

The relation between the parameters $x_i$ and $z_i$ can be found
at equilibrium, where $p_i=0$, using (\ref{spectralcurve}).
In the $SU(2)$ case, we have $N=2$ in (\ref{laxpair}) and (\ref{sigmazeta}).
It is therefore easy to compute
\bea
&&\det\left(
\begin{array}{cc}
\lambda-p_1& -m\Phi((x_1-x_2,\varpi)\\ -m\Phi(x_2-x_1,\varpi) & \lambda-p_2
\end{array}\right)=-m^2\Phi(x_1-x_2,\varpi)\Phi(x_2-x_1,\varpi)+\lambda^2\nonumber\\
&&+p_1p_2-\lambda(p_1+p_2)=
m^2h_1^\prime(\varpi)-m^2h_1^\prime(x_1-x_2)+\lambda^2-\lambda(p_1+p_2)+p_1p_2
\label{laxmatrixn=2}
\eea
In the last equality of the above equation we
used the following identity for the odd function $\sigma$ \cite{Erdelyi}
\be
\sigma(\lambda-(x_1-x_2))\sigma(\lambda+(x_1-x_2))=-\sigma^2(\lambda)\sigma^2(x_1-x_2)
[\wp(\lambda)-\wp(x_1-x_2)] \label{laxmatrixn=3}
\ee
and the explicit form of the Weierstrass function given in the Appendix
, eqs.(\ref{wei}) and (\ref{wei1}).

Then comparing with
$f(\lambda + m h_1(\varpi),\varpi)= m^2 h_1^\prime(\varpi)-z_1^2+\lambda^2$ one can see that
the dependence on the spectral parameter $\varpi$ cancels. This is a
general feature which is valid for arbitrary values of $N$. The
general relation between the parameters $z_i$ and the momenta and
positions of the Calogero-Moser system was found in
\cite{D'Hoker:1998xd}.
The vacua of the $\cN=1^*$ theory , whose quantum
potential is given by the Weierstrass $\wp$-function
\cite{Dorey:1999sj}, are given by the equilibrium
positions of the associated Calogero-Moser system. Solving the
equation $\wp^\prime(x_1-x_2)=0$ \cite{Dorey:1999sj} we then find
\footnote{We scale by a factor of minus four
the result found for the confining vacuum in \cite{Dorey:1999sj}:
$\wp(x_1-x_2)\vert_{x_i=x_{conf}}=N^2/24(E_2(q)-1/NE_2(q^{1/N}))$
, due to the different choice of the semi-period $\omega_1=i\pi$
used in that paper.}
\bea
z_1^2&=&m^2h_1^\prime(x_1-x_2)\vert_{x_i=x_{vac}}= -m^2
\pi^2 e_i -\frac{m^2\pi^2}{3}E_2(q)
\eea
in agreement with (\ref{z1s}) after the replacement $im_c\to m$.

\section{Summary of Results}

Here we summarize our results. We consider ${\cal N}=1$ gauge theories with
or without adjoint matter. The two main examples are mass deformations or
pure ${\cal N}=2$ and ${\cal N}=2^*$ gauge theories
down to ${\cal N}=1$. The later one is refereed as ${\cal N}=1^*$.
For the chiral correlators and gaugino
condensates one finds
\bea
\langle {\rm tr}\, \varphi^J\, \rangle_{{\cal N}=1}
&=&\langle {\rm tr}\, \varphi^J e^{-\int_{\C^2} \alpha_{(2,0)}\wedge {\rm tr}\,  W({\cal F})}
\, \rangle_{{\cal N}=2}=\langle {\rm tr}\, \varphi^J\, \rangle_{{\cal N}=2}\nn\\
\langle \tr \lambda\lambda\vphi^J \rangle_{\cN=1} &=& - \ft{2m}{(J+2)(J+1)}
  \,q\, \partial_q \,
 \langle {\rm tr}\, \varphi^{J+2} \rangle_{{\cal N}=2}
 \eea
where the r.h.s. correlators are to be evaluated in the parent ${\cal N}=2$
or ${\cal N}=2^*$ SYM
and then their v.e.v.'s have to be chosen so to minimize the quantum potential.
For instance for the ${\cal N}=1$ descending for pure ${\cal N}=2$
gauge theory with gauge group $SU(2)$
in the $SU(2)$ unbroken phase one finds
$u_{1,2}=\langle {\rm tr}\, \varphi^2\, \rangle=\pm 4 \Lambda^2$ and higher
order scalar correlators are given by the (\ref{su2ringrelations}).

 For ${\cal N}=1^*$ with gauge group $SU(2)$ one finds
\begin{eqnarray}
u_1&=&\frac{m^2}{6}( 4\, E_2(q^2)-1)\nn \\
u_2&=&\frac{m^2}{6}(E_2(-\sqrt{q})-1)\nn \\
u_3&=&\frac{m^2}{6}(E_2(\sqrt{q})-1)
\end{eqnarray}
$u_1$ gives the result in
the Higgs phase, while in the confining phase $u=\ft12(u_{2}-u_{3})$,  
with $u_{2,3}$ the contributions coming from the two vacua.
The results for pure ${\cal N}=1$ gauge theory can be read from this
by sending $m\to \infty$ keeping $\Lambda^2_{{\cal N}=1}=m^2 q^{1\over 2}$
in the confining phase:
$u_{2,3}=\pm 4\Lambda^2_{{\cal N}=1}$.

The scalar correlator with the next higher power is given by
\bea
\langle\tr\varphi^4\rangle
&=&\ft12 u^2
+\ft13 {m}^2 (E_2(q)-1)u+\ft{1}{18}{m}^4
\left(\ft15+E_2(q)-E_2(q)^2-\ft15 E_4(q)\right)
\eea
In  a similar way one can study gravitational corrections.  For instance
for ${\cal N}=1^*$ with gauge group $U(1)$ one finds
\bea
\langle \tr \vphi^2 \rangle_{{\cal N}=1^*}&=&   (m^2-\hbar^2)\,(1+q+3 q^2+4 q^3+7 q^4+\ldots)
\nn\\
&=&(m^2-\hbar^2)\,\sum_{d|k} d \,q^k =
-\ft{1}{24}\,(m^2-\hbar^2)\, E_2(q)
\label{fin}\eea
The term $\hbar^2$ represents the only gravitational correction to $u$!
We remind the reader that even if our focus in this paper
was on gauge theories and, in order to keep our formulae down
to a manageable size, we have omitted gravitational corrections,
they are easily accounted for and could be compared with analogous results
obtained for the topological string \cite{Klemm:2002pa,Hollowood:2003cv,Eguchi:2003sj}.
The $U(N)$ case of (\ref{fin}) can be easily extracted from \cite{Flume:2002az}.

\section*{Acknowledgements}
We thank U. Bruzzo for useful discussions.
R.P. would like to thank I.N.F.N. for supporting a visit to the
University of Rome II, "Tor Vergata". This work was supported in
part by the EC contract MRTN-CT-2004-512194, the MIUR-COFIN contract
2003-023852, the NATO contract PST.CLG.978785, the INTAS contracts
03-51-6346 and the Volkswagen foundation of Germany.

\begin{appendix}

\section{Elliptic functions}
\setcounter{equation}{0}
Here we collect some useful formulae and definitions:

Theta functions \be \vartheta_{
  \alpha \beta}(v|q)=\sum_{n\in \Z}\, q^{{1\over 2}(n-{\alpha\over 2})^2} \, e^{2\pi
i(n-{\alpha\over 2})(v-{\beta\over 2})} \ee with $q=e^{2\pi i
\tau}$. We adopt the standard shorthand notation:
$\vartheta_1=\vartheta_{11}$, $\vartheta_2=\vartheta_{10}$,
$\vartheta_3=\vartheta_{00}$, $\vartheta_4=\vartheta_{01}$. At
$v=0$ the theta functions satisfy the Riemann Identity \be
\vartheta_3(0|q)^4-\vartheta_2(0|q)^4-\vartheta_4(0|q)^4=0 \ee

Einstein series
\bea
E_2(q)&=&1-24 f_1(q)\nn\\
E_4(q)&=&1+240 f_3(q)\nn\\
E_6(q)&=&1-504 f_5(q)
\eea
with
\be
f_p(q)=\sum_{n=1}^\infty {n^p \, q^n\over 1-q^n}\label{fps}
\ee
 Weierstrass Function
\be \wp(z,w_1,w_2)=-\frac{1}{3}\left(\frac{\pi}{
2\omega_1}\right)^2 E_2(q)-\,\partial^2_z \,\log
\vartheta_1\left(\ft{z}{2\omega_1}|q\right)
\label{wei}
\ee
with $\tau=\omega_1/\omega_2$ and $\omega_{1,2}$ the
half-periods. Typical choices for the periods are
$(\omega_1,\omega_2)=(\ft12,\ft{\tau}{2})$ or
$(\omega_1,\omega_2)=(\ft{\pi}{2},\ft{\pi\tau}{2})$. The two notations are
related by $\wp(tz\vert t\omega_1,t\omega_2)=t^{-2}\wp(z\vert
\omega_1,\omega_2)$. In particular one finds
$\wp(\ft{\pi}{2}\vert\ft{\pi}{2},\ft{\pi \tau}{2})=e_1$ and
$\wp(\ft{1}{2}\vert\ft{1}{2},\ft{ \tau}{2})=\pi^2 e_1$. In the main text we use
the shorthand notation
\be
\wp(z)\equiv \wp(z\vert\ft{1}{2},\ft{
\tau}{2})
\label{wei1}
\ee
  The complete elliptic integral of first
kind \be K(k^2)\equiv \int_0^1 {dt\over (1-t^2)(1-k^2\, t^2)}
\label{kint} \ee
For the reader's convenience we collect here the $q$-expansions
of the $e_i$'s given in (\ref{eis})
 \bea
e_1&=&\frac{2}{3} + 16\,q + 16\,q^2 +
  64\,q^3 + 16\,q^4 + 96\,q^5 +
  64\,q^6 + 128\,q^7 +
  16\,q^8+\ldots\nonumber\\
e_2&=&-\frac{1}{3}-8\,q^{1/2}-8q
  - 32\,q^{3/2} - 8\,q^2 - 48\,q^{5/2} -
  32\,q^3 - 64\,q^{7/2} - 8\,q^4 -
  104\,q^{9/2} - 48\,q^5 \nonumber\\&&- 96\,q^{11/2} -
  32\,q^{6} - 112\,q^{13/2} -
  64\,q^{7} - 192\,q^{15/2} - 8\,q^{8}+\ldots\nonumber\\
e_3&=& -\frac{1}{3} + 8 q^{1/2} - 8 q + 32\ q^{3/2}-8 q^2 + 48
q^{5/2} - 32 q^3 +
    64 q^{7/2} - 8 q^4 + 104 q^{9/2} - 48 q^5\nonumber\\&& + 96 q^{11/2} - 32 q^6 +
    112 q^{13/2} - 64 q^7 + 192 q^{15/2} - 8 q^8+\ldots
\label{eisexp}\eea
and the following identities
\bea
e_1&=& -\frac{2}{3} \left(E_2(q) - 2 E_2(q^2)\right)
\nonumber\\
e_2&=& -\frac{2}{3} \left(E_2(q) - \frac{1}{2} E_2(\sqrt{q})\right)
\nonumber\\
e_3&=& -\frac{2}{3} \left(E_2(q) - \frac{1}{2} E_2(-\sqrt{q})\right)
\label{e-id}
\eea

\end{appendix}

\end{document}